\documentclass{ws-procs9x6}

\begin{document}

\title{LIFE ON EARTH -- AN ACCIDENT?\\
Chiral Symmetry and the Anthropic Principle}

\author{Ulf-G. MEI{\ss}NER$^*$ [for the NLEFT Collaboration]}

\address{HISKP and BCTP, Bonn University,
D-53115 Bonn, Germany\\
IAS, IKP and JCHP, Forschungszentrum J\"ulich
D-52425 J\"ulich, Germany\\
$^*$E-mail: meissner@hiskp.uni-bonn.de\\
www.itkp.uni-bonn.de/$\tilde{\,\,\,}$meissner}

\begin{abstract}
I discuss the fine-tuning of the nuclear forces and in the formation of
nuclei in the production of the elements in the Big Bang and in
stars. 
\end{abstract}


\bodymatter

\section{Definition of the problem}
\label{sec:intro}

The elements that are pertinent to life on Earth are generated in the 
Big Bang and in stars through the fusion of protons, neutrons and nuclei.
In Big Bang nucleosynthesis (BBN), alpha particles ($^4$He nuclei) and some
heavier elements are generated. Life essential elements like $^{12}$C and
$^{16}$O are generated in hot, old stars, where the so-called triple-alpha
reaction plays an important role. Here, two alphas fuse to produce the
instable, but long-lived $^8$Be nucleus. As the density of $^4$He nuclei in
such stars is high, a third alpha fuses with this nucleus before it decays.
However, to generate a sufficient amount of $^{12}$C and $^{16}$O, an excited
state in $^{12}$C at an excitation energy of 7.65~MeV with spin zero and
positive parity is required as pointed out by Hoyle long ago \cite{Hoyle}.
In a further step, carbon is turned into oxygen without such a resonant
condition. So we are faced with a multitude of fine-tunings which need to
be explained. We know that  all strongly interacting composites like 
hadrons and nuclei must emerge from the underlying gauge theory of the
strong interactions, Quantum
Chromodynamics (QCD), that is formulated in terms of quarks and gluons.
These fundamental matter and force fields are, however, confined. Further,
the mass of the light quarks relevant for nuclear physics is very small
and thus plays little role in the total mass of nucleons and nuclei. Finally,
protons and neutrons form nuclei. This requires the inclusion of
electromagnetism, characterized by the fine-structure constant $\alpha_{\rm
  EM}\simeq 1/137$. So the question we want to address in the following is:
How sensitive are these strongly interacting composites to variations in the
fundamental parameters of QCD+QED? or stated differently: how accidental is
life on Earth?

\section{The nuclear force at varying quark mass}
\label{sec:forces}

\begin{figure}[t]
\begin{center}
\psfig{file=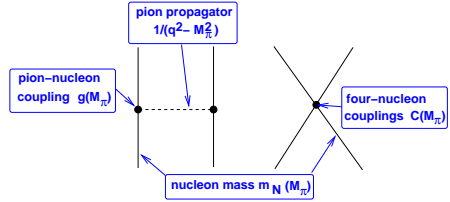,width=2.3in}
\end{center}
\caption{Explicit and implicit pion (quark) mass dependence of the
leading order nucleon-nucleon (NN) potential. Solid (dashed) lines denote
nucleons (pions).}
\label{fig:mpi}
\end{figure}

\begin{figure}[t]
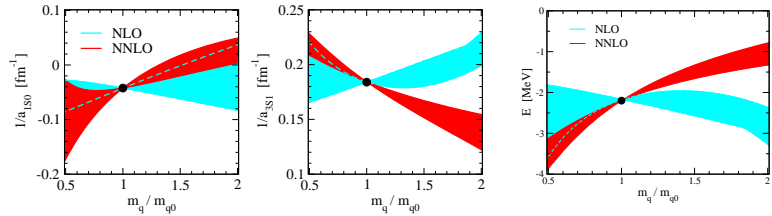

~~~~~\psfig{file=InvScatt_RunTPE.eps,width=2.5in}

\vspace{-2.8cm}

\hfill\psfig{file=BE_RunTPE.eps,width=1.305in}~~~~~~
\caption{Quark mass dependence of the inverse scattering length $1/a_{1S0}$
and $1/a_{3S1}$ and the deuteron binding energy. Here, $m_{q0}$ denotes the
physical light quark mass.}
\label{fig:run}
\end{figure}

Nuclear forces are best described by utilizing chiral effective field
theory (EFT) as pioneered by Weinberg \cite{Weinberg:1990rz}. The forces
between two, three and four nucleons are given by pion-exchange contributions
and short-distance multi-nucleon operators, the latter being accompanied
by low-energy constants that must be determined by a fit to data. For a
review, see Ref.~\cite{Epelbaum:2008ga}$\,$. In this scheme, the quark mass
dependence of the forces is generated explicitely (pion propagator) and
implicitly (pion-nucleon coupling, nucleon mass, 4N couplings), see 
Fig.\ref{fig:mpi}.
Throughout, we use the Gell-Mann--Oakes--Renner relation, $M_\pi^2 \sim
(m_u+m_d)$, so one can use pion and quark mass dependence synonymously.
 For any observable ${\cal O}$ of a hadron $H$, we can define
its quark mass dependence in terms of the so-called $K$-factor, 
$\delta {\cal O}_H/\delta m_f \equiv K_H^f \,({\cal O}_H/m_f)$, with $f=u,d,s$,
and $m_f$ the corresponding quark mass.
The pion mass dependence of pion and nucleon properties can be obtained from
lattice QCD combined with chiral perturbation theory as detailed in
Ref.\cite{Berengut:2013nh}$\,$. The pertinent results are: 
$K_{M_\pi}^q = 0.494^{+0.009}_{-0.013}$, $K_{F_\pi}^q = 0.048\pm 0.012$, and
$K_{m_N}^q = 0.048^{+0.002}_{-0.006}$, where $q$ denotes the average light
quark mass. For the quark mass dependence of the short-distance terms, 
one has to resort to modeling using resonance saturation\cite{Epelbaum:2001fm}. This induces
a sizeable uncertainty that might be overcome by lattice simulations in the
future. For the NN scattering lengths, this leads to $K^q_{1S0} = 2.3^{+1.9}_{-1.8}$,
$K^q_{3S1} = 0.32^{+0.17}_{-0.18}$ and $K^q_{\rm BE(deut)} =
-0.86^{+0.45}_{-0.50}$ (with BE denoting the binding energy), extending and 
improving earlier work based on EFTs and 
models\cite{Muther:1987sr,Beane:2002vs,Epelbaum:2002gb,Flambaum:2007mj,Soto:2011tb}.
The running of the NN scattering lengths and the deuteron BE with the light quark 
mass is shown in Fig.~\ref{fig:run}.
In addition to shifts in $m_q^{}$, we shall also consider the
effects of shifts in $\alpha_{\rm EM}^{}$. The
treatment of the Coulomb interaction in the nuclear lattice EFT framework is
described in detail in Ref.~\cite{Epelbaum:2010xt}.
\section{Constraints from Big Bang Nucleosynthesis}
\label{sec:bbn}

With the results from the previous section, one can now analyze what
constraints the element abundances in BBN on possible quark mass variations imply.
To answer this question, we also need the variation of $^3$He and $^4$He
with the pion mass. Following Ref.~\cite{Bedaque:2010hr} (BLP), these can be
obtained by convoluting the 2N $K$-factors with the variation of the
3- and 4-particle BEs with respect to the singlet and triplet NN scattering 
lengths. This gives $K_{^3{\rm He}}^q = -0.94\pm 0.75$ and  $K_{^3{\rm He}}^q 
= -0.55\pm 0.42$\cite{Berengut:2013nh}, which is consistent with a direct
calculation using nuclear lattice simulations, $K_{^3{\rm He}}^q = -0.19\pm
0.25$ and $K_{^3{\rm He}}^q = -0.16\pm 0.26$\cite{Lahde}. With this input,
we can calculate the BBN response matrix of the primordial abundances $Y_a$ at 
fixed baryon-to-photon ratio, $\delta Y_a/\delta m_q = \sum_{X_i} (\delta \ln
Y_a/ \delta \ln X_i)\, K_{X_i}^q$, with $X_i$ the relevant BEs for $^2$H,
$^3$H, $^3$He, $^4$He, $^6$Li, $^7$Li and  $^7$BE and the singlet NN
scattering length, using the updated Kawano code (for details,
see Ref.\cite{Berengut:2009js}). Combining the calculated with the observed
abundances, one finds that the most stringent limits arise from the
deuteron abundance [deut/H] and the $^4$He abundance normalized to the one
of protons, $^4$He($Y_p$), as most neutrons end up in the alpha
nucleus. Combining these leads to the constraint {$\delta m_q/m_q = (2\pm 4)\%$}.
In contrast to most earlier determinations, we provide reliable error
estimates due to the underlying EFT. However, as pointed out by BLP, one
can obtain an even stronger bound due to the neutron lifetime, which 
strongly affects  $^4$He($Y_p$). We have re-evaluated this constraint under
the model-independent assumption that {\em all} quark and lepton masses
vary with the Higgs vacuum expectation value $v$, leading to 
\begin{equation}
|\delta v/ v | = |\delta m_q/ m_q| \leq 0.9\%~. 
\end{equation}

\section{The fate of carbon-based life as a function of the fundamental parameters of QCD+QED}
\label{sec:fate}

\begin{figure}
\begin{center}
\psfig{file=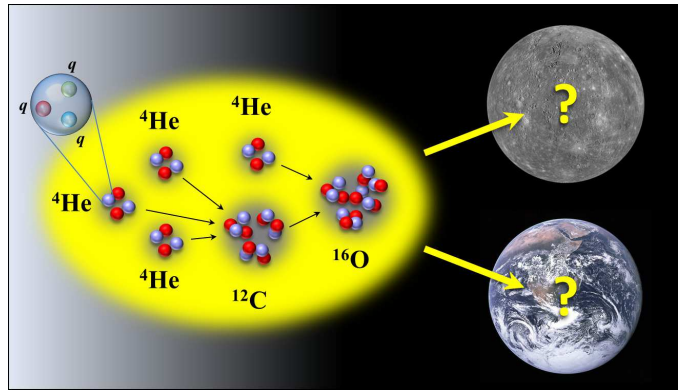,width=3.5in}
\end{center}
\caption{Graphical representation of the question of how fine-tuned is
life on Earth under variations of the average light quark mass and $\alpha_{\rm EM}$. 
Figure courtesy of Dean Lee.}
\label{fig:fate}
\end{figure}

\begin{figure}[t]
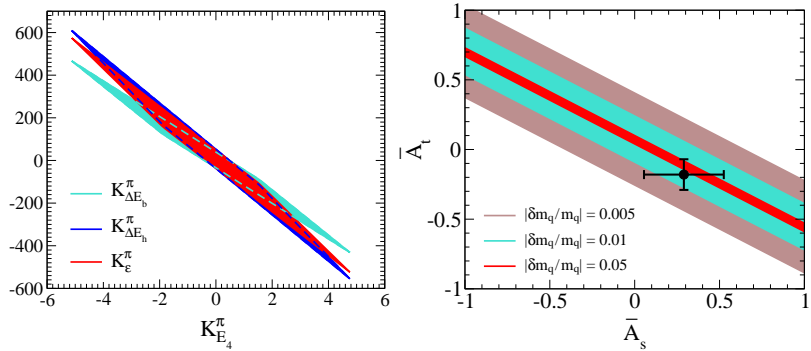

\begin{center}
\psfig{file=corelDeltaBEv5.eps,width=2.0in}
\psfig{file=hoyle_band_new.eps,width=2.15in}
\end{center}
\caption{Left panel: Sensitivities of $\Delta E_h^{}$, $\Delta E_b^{}$ and $\varepsilon$ 
to changes in $M_\pi^{}$, as a function of $K_{E_4^{}}^\pi$ under independent 
variation of $\bar A_s^{}$ and $\bar A_t^{}$ over the range $\{-1 \ldots 1\}$.   
The bands correspond to $\Delta E_b^{}$, 
$\varepsilon$ and  $\Delta E_h^{}$ in clockwise order.
Right panel: 
``Survivability bands'' for carbon-oxygen based life from
  Eq.~(\ref{final_res}), due to  $0.5\%$ (broad outer band), $1\%$ (medium
  band) and $5\%$ (narrow inner band) changes in $m_q^{}$ in terms of the
  input parameters $\bar A_s^{}$ and $\bar A_t^{}$. The most up-to-date 
  N$^2$LO analysis of $\bar A_s^{}$ and $\bar A_t^{}$ from
  Ref.\cite{Berengut:2009js} is given by the data point with 
  horizontal and vertical error bars.}
\label{fig:band}
\end{figure}

I now turn to the central topic of this talk, namely how fine-tuned is the
production of carbon and oxygen with respect to changes in the fundamental
parameters of QCD+QED? Or, stated differently, how much can we detune these
parameters from their physical values to still have an habitable Earth as shown
in Fig.~\ref{fig:fate}. To be more precise, we must specify which parameters
we can vary. In QCD, the strong coupling constant is tied to the nucleon mass
through dimensional transmutation. However, the light quark mass (here, only
the strong isospin limit is relevant) is an external parameter. Naively, one
could argue that due to the small contribution of the quark masses to the
proton and the neutron mass, one could allow for sizeable variations. However,
the relevant scale to be compared to here is the average binding energy per
nucleon, $E/A \leq 8\,$MeV (which is much smaller than the nucleon mass). 
As noted before, the Coulomb repulsion between protons is an important
ingredient in nuclear binding, therefore we must also consider changes in
$\alpha_{\rm EM}$. The tool to do this are nuclear lattice simulations, which
allowed e.g. for the first {\em ab initio} calculation of the Hoyle 
state\cite{Epelbaum:2011md}. Let us consider first QCD (for details, see
Refs.\cite{Epelbaum:2012iu,Epelbaum:2013wla}). We want to calculate
the variations of the pertinent energy differences  in the
triple-alpha process $\delta \Delta E/ \delta M_\pi$, which according to
Fig.~\ref{fig:mpi} boils down to (we consider small variations around
the physical value of the pion mass $M_\pi^\mathrm{ph}$):
\begin{eqnarray}
\left. \frac{\partial E_i^{}}{\partial M_\pi^{}} \right|_{M_\pi^\mathrm{ph}} &= &
\left. \frac{\partial E_i^{}}{\partial \tilde M_\pi^{}} \right|_{M_\pi^\mathrm{ph}}
+ x_1^{} \left. \frac{\partial E_i^{}}{\partial m_N^{}} \right|_{m_N^\mathrm{ph}}
+ x_2^{} \left. \frac{\partial E_i^{}}{\partial \tilde g_{\pi N}^{}} \right|_{\tilde g_{\pi N}^\mathrm{ph}} 
\nonumber \\
& +& x_3^{} \left. \frac{\partial E_i^{}}{\partial C_0^{}}
\right|_{C_0^\mathrm{ph}}
+ x_4^{} \left. \frac{\partial E_i^{}}{\partial C_I^{}} \right|_{C_I^\mathrm{ph}},
\label{Eeq2}
\end{eqnarray}
with the definitions
\begin{align}
& x_1^{} \equiv \left. \frac{\partial m_N^{}}{\partial M_\pi^{}} 
\right|_{M_\pi^\mathrm{ph}}, ~~
x_2^{} \left. \equiv \frac{\partial g_{\pi N}^{}}{\partial M_\pi^{}}
\right|_{M_\pi^\mathrm{ph}}~
x_3^{} \equiv \left. \frac{\partial C_0^{}}{\partial M_\pi^{}}
\right|_{M_\pi^\mathrm{ph}}, ~~
x_4^{} \equiv \left. \frac{\partial C_I^{}}{\partial M_\pi^{}} \right|_{M_\pi^\mathrm{ph}},
\label{xy}
\end{align}
with $\tilde{M_\pi}$ the pion mass appearing in the pion-exchange potentials
The various derivatives in Eq.~(\ref{Eeq2}) can be obtained precisely using
Auxiliary Field Quantum Monte Carlo techniques and the $x_i$ ($i = 1,2,3,4$) are related to
the pion and nucleon $K$-factors determined in Sec.~\ref{sec:forces}. 
The scheme-dependent quantities $x_{3,4}$ can be traded for the
pion-mass dependence of the inverse singlet and triplet scattering lengths,
$\bar A_{s}^{} \equiv {\partial a_{s}^{-1}}/{\partial M_\pi^{}}|_{M_\pi^{\rm
    ph}}$,  $\bar A_{t}^{} \equiv {\partial a_{t}^{-1}}/{\partial M_\pi^{}}|_{M_\pi^{\rm ph}}$.
We can then
express all energy differences appearing in the triple-alpha process 
($\Delta E_b^{} \equiv E_8^{} - 2 E_4^{}, \Delta E_h^{} \equiv E_{12}^\star - E_8^{} - E_4^{},
\varepsilon =  E_{12}^\star - 3E_4^{}$, with $E_4^{}$ and $E_8^{}$ for the energies of 
the ground states of $^{4}$He and $^{8}$Be, respectively, and $E_{12}^\star$ denotes the
energy of the Hoyle state) as functions
of $\bar A_{s}$ and $\bar A_{t}$. One finds that all these energy differences
are correlated, i.e. the various fine-tunings in the triple-alpha process are not independent
of each others, see the left panel of Fig.~\ref{fig:band}. Further, one finds a 
strong dependence on the variations of the
$^4$He BE, which is strongly suggestive of the $\alpha$-cluster structure of
the $^8$Be, $^{12}$C and Hoyle states.  Such correlations related to the production 
of carbon have indeed been speculated upon earlier~\cite{Livio,WeinbergFacing}.
Consider now the reaction rate of the triple-alpha process as given by
$r_{3 \alpha}^{} \sim N_\alpha^3 \Gamma_\gamma^{} \exp \left( -{\varepsilon}/{k_{\rm B}^{} T} \right)$,
with $N_\alpha$ the $\alpha$-particle number density in the stellar plasma with temperature $T$, 
$\Gamma_\gamma = 3.7(5)\,{\rm meV}$ the radiative width of the Hoyle state and 
$k_B$ is Boltzmann's constant.
The stellar modeling calculations of Refs.~\cite{Oberhummer:2000mn,Oberhummer-astro} suggest 
that sufficient abundances of both carbon and oxygen can be maintained within an envelope of 
$\pm 100$~keV around the empirical value of $\varepsilon = 379.47(18)$~keV. 
This condition can be turned into a constraint on shifts in $m_q^{}$ that reads
(for more details, see Ref.\cite{Epelbaum:2013wla})
\begin{equation}
\left| \Big[ 0.572(19) \, \bar A_s^{} + 0.933(15) \, \bar A_t^{} 
- 0.064(6)  \Big]  
\left(\frac{\delta m_q^{}}{m_q^{}} \right) \right|  < 0.15\%~.
\label{final_res}
\end{equation}
The resulting constraints on the values of $\bar
A_s^{}$ and $\bar A_t^{}$ compatible
with the condition $| \delta \varepsilon | < 100$~keV are visualized in the
right panel of Fig.~\ref{fig:band}.  The various shaded bands in this figure cover the 
values of $\bar A_s^{}$ and $\bar A_t^{}$ consistent
with carbon-oxygen based life, when $m_q^{}$ is varied by $0.5$\%, $1$\% and $5$\%.
Given the current theoretical 
uncertainty in $\bar A_s^{}$ and $\bar A_t^{}$, our results remain compatible 
with a vanishing $\partial \varepsilon / \partial M_\pi^{}$, in other words
with a complete lack of fine-tuning. Interestingly, Fig.~\ref{fig:band} (right panel) 
also  indicates that the triple-alpha process is unlikely to be fine-tuned
to a higher degree than $\simeq 0.8$\% under variation of $m_q^{}$. 
The central values of $\bar A_s^{}$ and $\bar A_t^{}$ from
Ref.\cite{Berengut:2009js} suggest that variations in the light quark masses 
of up to $2 - 3$\% are unlikely to be catastrophic to the formation of 
life-essential carbon and oxygen. A similar calculation of the tolerance for 
shifts in the fine-structure constant $\alpha_{\rm EM}^{}$
suggests that carbon-oxygen based life can withstand shifts of 
$\simeq 2.5$\% in $\alpha_{\rm EM}^{}$.

\section{A short discussion of the anthropic principle}
\label{sec:ant}

The Hoyle state dramatically increases the reaction rate of the
triple-alpha process. The resulting enhancement
is also  sensitive to the exact value of $\varepsilon$, which is
therefore the principal control parameter of
this reaction. As the Hoyle state is crucial to the formation of elements
essential to life as we know it, this state
has been nicknamed the ``level of life''~\cite{Linde}. 
Thus, the Hoyle state is often viewed as a prime manifestation of the anthropic
principle, which states that the observable 
values of the fundamental physical and cosmological parameters are restricted
by the requirement that life can form to 
determine them, and that the Universe be old enough for that to
occur~\cite{Carter,Carr}. See, however, Ref.~\cite{Kragh} for a thorough historical
discussion of the Hoyle state in view of the anthropic principle.
We remark that in the context of cosmology and string 
theory, the anthropic principle and its consequences have had a significant influence,
as reviewed recently in \cite{Schellekens:2013bpa}.
As noted already in Ref.\cite{WeinbergFacing}, the allowed variations in $\varepsilon$
are not that small, as $|\delta \varepsilon/\varepsilon | \simeq 25\%$ still allows
for carbon-oxygen based life. So one might argue that the anthropic principle is
indeed {\it not} needed to explain the fine-tunings in the triple-alpha process.
However, as we just showed, this translates into allowed quark mass variations
of $2-3\%$ and modifications of the fine-structure constant of about 2.5\%.
The  fine-tuning in the fundamental parameters is thus much more severe than the
one in the energy difference $\varepsilon$. Therefore, beyond such relatively small changes 
in the fundamental parameters, the anthropic principle indeed appears necessary to 
explain the observed abundances of $^{12}$C and $^{16}$O. 

\section{Summary and outlook}
\label{sec:sum}

In this talk, I have summarized recent developments in our understanding of
the fine-tuning in the generation of the life-essential elements as well as
the light elements generated in BBN. As shown, the allowed parameter
variations in QCD+QED are small, giving some credit to the anthropic
principle. To sharpen these conclusions, future work is required. On one side,
lattice QCD at sufficiently small quark masses will eventually be able to
give tighter constraints on the parameters $\bar A_{s,t}$ and on the other
side, nuclear lattice simulations have to be made more precise to reduce the
theoretical error in the binding and excitation energies and to provide {\it
ab initio} calculations of nuclear reactions, for first steps, 
see Refs.\cite{Rupak:2013aue,Pine:2013zja}$\,$.

\subsection*{Acknowledgments}

I would like to thank my NLEFT collaborators Evgeny Epelbaum, Hermann Krebs, 
Timo L\"ahde and Dean Lee for a most enjoyable collaboration.  I also
thank the organizers for their perfect job.
Work supported in part by DFG
and NSFC (Sino-German CRC 110), Helmholtz Association 
(contract VH-VI-417), BMBF\ (grant 05P12PDFTE), and the EU (HadronPhysics3 project)
Computational resources provided 
by the J\"{u}lich Supercomputing Centre (JSC) at the Forschungszentrum 
J\"{u}lich  and by RWTH Aachen.

\end{document}